# Re-appraisal and extension of the Gratton-Vargas two-dimensional analytical snowplow model of plasma focus evolution in the context of contemporary research

S K H Auluck

Abstract

Recent resurgence of interest in applications of dense plasma focus and doubts about the conventional view of dense plasma focus as a purely irrotational compressive flow have re-opened questions concerning device optimization. In this context, this paper re-appraises and extends the analytical snowplow model of plasma focus sheath evolution developed by F. Gratton and J.M. Vargas (GV) (Energy Storage, Compression and Switching, Ed. V. Nardi, H. Sahlin, and W. H. Bostick, Eds., vol. 2. New York: Plenum, 1983, p. 353) and shows its relevance to contemporary research. The GV model enables construction of a special orthogonal coordinate system in which the plasma flow problem can be simplified and a model of sheath structure can be formulated. The Lawrenceville Plasma Physics (LPP) plasma focus facility, which reports neutron yield better than global scaling law, is shown to be operating closer to an optimum operating point of the GV model as compared with PF-1000.

## I. Introduction:

One of the major insights gained during five decades of Dense Plasma Focus (DPF) research concerns the existence of some parameters of experimentally optimized DPF devices which remain approximately constant for variation in stored energy over 8 decades [1]. This compilation and analysis of data suggests that many gross features of plasma evolution in DPF are insensitive to details of plasma behavior on length and time scales smaller than those associated with the plasma current sheath (PCS). It should then be possible to construct theoretical models, such as the Lee model [2], for predicting these gross features using *oversimplified* physical assumptions – assumptions which are *known* to be coarse approximations of reality. The importance of such oversimplification is two-fold: the ability of models like the Lee model to reproduce some experimental results, typically the current profile, with a small number of adjustable parameters indicates *independence of global dynamics from local dynamics* [3] or alternatively, insensitivity to the coarseness of model approximations (such as representing a 3-dimensional phenomenon by a zero-dimensional model). At the same time, such oversimplified models cannot predict some of the more exotic observations such as the toroidal directionality of energetic ions participating in the fusion reactions [4] or occurrence of axial magnetic field [5], which is shown to be related with generation of toroidal plasma structures [6] which apparently play some part in the fusion reactions [7].

This suggests that the physics of the DPF needs to be described at two levels:

1. at the level of global behavior (plasma shape, motion, current profile, inductance variation, conversion of energy stored in the capacitor bank into kinetic, thermal and magnetic energy) and its relation with device parameters (shapes and dimensions of electrodes, location and shape of the initial plasma, nature and pressure of fill gas, capacitance, inductance, resistance and charging voltage of the capacitor bank) with simplest possible physical assumptions;
2. at the level of internal phenomena of a plasma structure whose gross behavior is already known from level 1.

Such approach would have significant implications. Recent reviews [8-11] suggest imminent emergence of DPF as a technology platform for commercially significant applications. At the same time, there is also some revival of interest in DPF as a device for generating fusion energy [12] and other nuclear reactions [13]. Device optimization issues are very different in the two cases. Ability to predict global dependencies between control parameters and quantitative performance criteria would be more relevant for applications to material science and other uses of DPF radiation. A simplified global model, such as the Lee model, would be a tool of choice for device optimization since it covers many different phases of plasma development and provides valuable insights which can guide the experimenter both in interpreting his results and in planning his future course of action.

On the other hand, the significantly-higher-than-thermal fusion reaction rate in DPF [14] must be adequately explained in all its properties, including the toroidal directionality of energetic ion motion inferred [4] from fusion reaction product diagnostics, for the purpose of exploring its potential as a fusion energy producer. Even sophisticated simulations [15, 16] do not address this issue. Some over-simplification of physics may be necessary to construct a tractable model with transparent assumptions, which yields useful, experimentally testable predictions for the exotic phenomena linked with the high fusion reactivity; such model can provide the starting point for subsequent, experimentally-driven refinements. Any model, which is *inherently incapable* of predicting non-trivial properties of fusion reactions in DPF would be unreliable for the purpose of predicting its scaling behavior in unexplored regions of the parameter space, particularly where designs of larger-than-state-of-art devices are being evaluated for practical realization.

Traditionally, the fusion reaction rate in the DPF has been conceptualized [17, 18] under the tacit assumption that the plasma evolution must be purely irrotational. However, recent results [5,6,7] as well as accumulated data over several decades of research [4] are at variance with this assumption. It has been recently suggested [4] that the peculiar circumstance of a fully ionized plasma magnetically driven into an initially neutral medium leads to a "constrained dynamics" situation, in which, the magnetic driving force is governed by external power source dynamics and resistivity-dependent dynamic skin effect, the pressure gradient force is governed by processes which add, subtract, redistribute, heat or cool particle species and the acceleration is governed by viscous drag of the neutrals. These processes are independent and very dissimilar; under these circumstances momentum balance cannot be satisfied unless the momentum conservation equation contains a term not constrained by any aspect of device operation, which instantaneously makes up the deficit in momentum balance. In azimuthally symmetric plasma, electron momentum convection by azimuthally streaming electrons can serve this role [4]. The corresponding ion counter-current would then explain many observed features of the neutron emission data.

The principle of constrained dynamics provides only a necessary condition: that the plasma acceleration should be governed by a process independent of the magnetic and thermal pressures, such as viscous drag by neutral gas through which the plasma is forced to move. This idea can be pursued further only if an analytic theory is able to calculate the distribution of the externally supplied current, density and temperature within the plasma so that the extent of momentum imbalance and its dependence on control parameters can be estimated.

This assumes importance in view of the recent interest [19] in the possibility of aneutronic fusion energy based on the p-$^{11}$B reaction using DPF [12]. The cross-section of this reaction peaks to 1.2 barn, more than 1/5$^{th}$ the peak cross-section of D-T reaction at a center-of-mass energy of 550 kev; occurrence of this order of ion energy has been inferred from the full width at half maxima of neutron time-of-flight spectra in *many* DPF and z-pinch devices [4]. The idea that this is caused by a *linear* beam of ions accelerated by induced axial electric field generated by instabilities or anomalous resistance is belied by evidence of accelerated ions in both forward and reverse axial directions as well as by evidence showing radially directed motion [4]. Clearly, an analytic theory, however approximate, which gives a global picture of dependencies between control parameters and properties of the accelerated ion population, would be very valuable for plotting the road map for the quest for aneutronic fusion energy [12] and other nuclear reactions [13].

In this context, the two-dimensional analytic snowplow model developed by Gratton and Vargas (GV) in 1970s [20] has an important, though largely un-appreciated, role to play. Unlike the Lee model, which is based on numerical solution of ordinary differential equations covering up to six phases of plasma development and which uses adjustable parameters to provide a *post facto* good fit to experimental current waveform, the GV model is based on *analytical solution* of partial differential equations, is limited to axial rundown and radial collapse phases only, is not valid in the pinch phase, uses no adjustable parameters and yields closed-form formulas for many cases. The Buenos Aires Group has compared this model with experiments [21] and has shown a reasonably good agreement, which can be further improved in principle by incorporating adjustable parameters in the spirit of the Lee model [2]. This approach has been used to estimate electrode erosion in a coaxial gun [22] used for material coating applications. In spite of some of its positive features, the GV model has not been extensively used by DPF community. This is probably because of relative inaccessibility of the GV work, (which was published only in conference proceedings), its weakness in quantitative predictions due to neglect of circuit resistance and because, unlike the Lee model, their work has not been available in a ready-to-use form. All these aspects are probably reflections of the fact that their work was far ahead of its time: it arrived much before the spread of personal computers and internet. On the other hand, GV model has led to significant insights into limiting factors affecting DPF performance [23,24,25] and also to semi-empirical guidelines for design of DPF devices [26].

The importance of the GV model is that it describes the global dynamics of DPF in a qualitative, (sometimes semi-quantitative), manner using an idealized analytical snowplow model with oversimplified physical assumptions, *indicating that details of local plasma dynamics do not affect the global behavior to any significant extent*. Many questions of practical relevance can therefore be addressed in a generic manner without recourse to sophisticated simulation. It certainly has some basic limitations rooted in its oversimplification, so that it may never provide a quantitative description of the DPF. At the same time, because of its analytical structure, it has the potential for providing a platform for construction of more detailed physical models.

Purely numerical models of physical phenomena necessarily have to make some simplifying assumptions, many times unstated, whose effect on the final physical picture is often obscured by the "black-box" of computational algorithms. Use of adjustable parameters to produce agreement with experimental numbers tacitly assumes that the particular set of numerical values of adjustable parameters, which maximizes agreement with experimental data, *is a unique set*: the possibility that a different group of researchers may come up with a different set which gives a comparable fit to the same data cannot be ruled out at the outset; later, the GV model is shown to provide a "good fit" to the same data for which Lee model also gives a good fit although for a different set of parameters, by choosing some ill-defined parameters judiciously. Also, it can never be claimed, without a large number of both experiments and computations, that the fitted parameters for one particular device will have constant values over a wide range of operating parameters such as voltage or pressure. These restrictions limit the credibility of performance projections based on numerical models with adjustable parameters for devices under development or planning.

Non-parametric analytical models work with a different paradigm. Instead of trying to predict numbers, they try to predict non-numerical, relational features such as trends and profiles. For example, the Gratton-Vargas model generates the characteristic shape of the DPF plasma and its current profile. The absence of fitted parameters reduces practical utility of non-parametric analytical models for predicting or analyzing experimental data. At the same time, it enhances the worth of their qualitative or semi-quantitative predictions because of their greater transparency.

Numerical and analytical models both represent approximations to the undiscovered physics, which play complementary roles on the path to discovery. For example, the fitted parameters of the Lee model 'encode' the unknown physics in some sense: a systematic empirical study of how the fitted parameters are related to device properties would give important clues to the underlying relationships between control parameters and device performance. On the other hand, study of the discrepancy between predictions of a non-parametric analytical model and corresponding experiment should also lead to enhancement of understanding about the role various physical phenomena play in governing device properties. A good illustration of the latter approach is found in the work of Bruzzone et.al. [22].

This paper aims to provide a re-appraisal of the work of Gratton and Vargas [20] and demonstrate its relevance to contemporary research in the altered context of widespread availability of cheap computing power and growing appreciation of commercial possibilities. The GV model is extended to the case of non-zero circuit resistance. It is shown that its analytical representation of plasma shape can be utilized to construct a special curvilinear coordinate system in which the plasma dynamics takes a simpler form; this can be developed into a theory of the sheath structure, although that development is beyond the scope of the present paper. The snowplow phenomenon may be thought of as a process of transferring energy from the capacitor bank to a dense magnetized plasma formation. The GV formalism then provides a unique perspective on the device optimization problem, which may be related to recent reports [12] of better–than–scaling–law neutron yield.

To fulfill this aim, some departure from the GV approach has been necessary:

- Only those results necessary for discussion in the context of contemporary research and sufficient to serve as a tutorial introduction are reviewed.
- Discussion of specific solutions is emphasized more than the general features of the formalism.
- SI units are used in place of CGS.
- Nomenclature is altered and assumptions are re-stated in a different form for the convenience of extension of the GV model.

This paper is not meant to be a substitute for the original papers, which contain much information not covered in this paper.

The next section revisits the GV model in a tutorial format with a view to facilitate its use in DPF research and extends it to the case of non-zero circuit resistance. Section III provides a discussion of the algorithmic implementation of the GV model and compares it with experimental data. Section IV describes outline of a model of sheath structure based on a special coordinate system constructed using the analytic representation of plasma shape in the GV model; some features of the coordinate system are described in the Appendix. Section V discusses some aspects of the device optimization problem using the GV model. Section VI presents summary and conclusions.

## II. Overview of the GV model:

The plasma current sheath (PCS) in the DPF, assumed to be azimuthally symmetric, can be visualized as a moving region in cylindrical (r,z) space, bounded by electrodes at appropriate radial coordinates and carrying a total current I(t), varying with time t, within two curved surfaces $\Sigma_1 \Rightarrow \psi_1(r,z,t) \equiv z - f_1(r,t) = 0$ and $\Sigma_2 \Rightarrow \psi_2(r,z,t) \equiv z - f_2(r,t) = 0$, which mark boundaries between current-carrying and current-free zones. The region around the front surface $\Sigma_1$ experiences a "wind pressure" because of ingestion of a flux of momentum density $(-\rho_0 v_n)(-v_n)$ into this region, where $-v_n$ is the component of "wind" velocity normal to the sheath surface, which moves with normal velocity $v_n$ with respect to the electrodes; $\rho_0$ is mass density of the fill gas. The region around the back surface $\Sigma_2$ experiences a magnetic pressure $B_\theta^2/2\mu_0$. The two dimensional analytical electromechanical model of Gratton and Vargas [20] assumes that at an imaginary mean surface $\Sigma \Rightarrow \psi(r,z,t) \equiv z - \bar{f}(r,t) = 0$ between the front and back surfaces, these two pressures are equal and this assumption, usually referred as the snowplow model[1], enables construction of a partial differential equation for this mean surface, referred henceforth as the GV surface. The starting point of the GV model is the snowplow model equation

$$\rho_0 v_n^2 = B_\theta^2/2\mu_0 \Rightarrow v_n = B_\theta/\sqrt{2\mu_0\rho_0} \ . \qquad 1$$

The unit vector $\hat{\xi}$ normal to the surface $\psi(r,z,t) = 0$, which makes angle $\phi$ with respect to the z-axis (illustrated in fig 1), is defined by

$$\hat{\xi} \equiv \vec{\nabla}\psi/\left|\vec{\nabla}\psi\right| \equiv (\sin\phi, 0, \cos\phi) \qquad 2$$

The total derivative of $\psi$ is zero: $\partial_t \psi + \vec{v}.\vec{\nabla}\psi = 0$, where $\vec{v}$ is the velocity of PCS. Therefore,

$$v_n = \left|\vec{v}.\hat{\xi}\right| = -\partial_t\psi/\left|\vec{\nabla}\psi\right| = -\partial_t\psi/\sqrt{(\partial_r\psi)^2 + (\partial_z\psi)^2} \ . \qquad 3$$

At the back surface, the azimuthal magnetic field is given by

$$B_\theta = \mu_0 I(t)/2\pi r \qquad 4$$

From 1, 3 and 4, the following partial differential equation is obtained

---

[1] The original snowplow model of Rosenbluth and Garwin for a toroidal pinch assumes that all mass swept by the current sheath piles up in a thin layer ahead of the magnetic piston. Because of the curved current sheath in the DPF, the plasma spills out tangentially similar to a snowplow piling snow on the sidewalks as its clears the road.

$$\partial_t \psi + \sqrt{(\partial_r \psi)^2 + (\partial_z \psi)^2} \frac{\mu_0 I(t)}{2\pi r \sqrt{2\mu_0 \rho_0}} = 0 \qquad 5$$

This equation has the important property of admitting a scaling (referred as the GV scaling, whose significance is discussed later) of coordinates and time: introducing a scale length 'a', (which may be the anode radius for the sake of convenience) to generate dimensionless coordinates $\tilde{r} \equiv r/a$ and $\tilde{z} \equiv z/a$ and defining a dimensionless parameter $\tau$ by the relation

$$\tau(t) = \frac{\mu_0}{\pi a^2 \sqrt{2\mu_0 \rho_0}} \int_0^t I(t') dt' \qquad 6$$

yields the scaled partial differential equation for the GV surface

$$\partial_\tau \psi + \sqrt{(\partial_{\tilde{r}} \psi)^2 + (\partial_{\tilde{z}} \psi)^2} \frac{1}{2\tilde{r}} = 0 \qquad 7$$

Note that unlike real time, the dimensionless parameter $\tau(t)$ is not a monotonically increasing variable; it reaches a maximum $\tau_{max}$ when the current reaches zero. The quantity $Q_m \equiv \pi a^2 \sqrt{2\mu_0 \rho_0}/\mu_0$ has dimensions of charge and GV refer to it as the 'mechanical equivalent of charge'. The parameter $\tau(t)$ is referred henceforth as '$\tau$ime'.

Equation 7 admits solutions separable in $\tau$: Putting $\psi(\tilde{r},\tilde{z};\tau) = \tau/2 + \psi_0(\tilde{r},\tilde{z})$ leads to

$$(\partial_{\tilde{r}} \psi_0)^2 + (\partial_{\tilde{z}} \psi_0)^2 = \tilde{r}^2 \qquad 8$$

A particular solution of 8 separable in coordinates can be constructed as follows: let $\psi_0(\tilde{r},\tilde{z}) = \psi_{\tilde{r}}(\tilde{r}) + \psi_{\tilde{z}}(\tilde{z})$. Then, 8 becomes

$$(d_{\tilde{z}} \psi_{\tilde{z}})^2 = \lambda^2; (d_{\tilde{r}} \psi_{\tilde{r}})^2 = \tilde{r}^2 - \lambda^2 \qquad 9$$

The resulting solution of 7

$$\psi(\tilde{r},\tilde{z};\tau) = \frac{\tau}{2} \pm \lambda\tilde{z} + C + \frac{1}{2}\frac{\tilde{r}}{\lambda}\sqrt{\frac{\tilde{r}^2}{\lambda^2}-1} - \frac{1}{2}\mathrm{ArcCosh}\left(\frac{\tilde{r}}{\lambda}\right) = 0 \qquad 10$$

shows a profile which propagates at the constant dimensionless 'velocity' $d\tilde{z}/d\tau$ equal to 1/2 while maintaining its shape.

More general solutions of 7 can be constructed by the method of characteristics [20]. Define generalized momenta $p_{\tilde{r}} \equiv \partial_{\tilde{r}}\psi$, $p_{\tilde{z}} \equiv \partial_{\tilde{z}}\psi$ and Hamiltonian $H = (2\tilde{r})^{-1}\sqrt{p_{\tilde{r}}^2 + p_{\tilde{z}}^2}$; then 7 takes the Hamilton-Jacobi form $p_{\tau} + H = 0$ for $p_{\tau} \equiv \partial_{\tau}\psi$.

The Hamiltonian equations are then

$$\frac{d\tilde{r}}{d\tau} = \frac{\partial H}{\partial p_{\tilde{r}}} = \frac{p_{\tilde{r}}}{2\tilde{r}\sqrt{p_{\tilde{r}}^2 + p_{\tilde{z}}^2}} = \frac{p_{\tilde{r}}}{4\tilde{r}^2 H} \qquad 11$$

$$\frac{d\tilde{z}}{d\tau} = \frac{\partial H}{\partial p_{\tilde{z}}} = \frac{p_{\tilde{z}}}{2\tilde{r}\sqrt{p_{\tilde{r}}^2 + p_{\tilde{z}}^2}} = \frac{p_{\tilde{z}}}{4\tilde{r}^2 H} \qquad 12$$

$$\frac{dp_{\tilde{r}}}{d\tau} = -\frac{\partial H}{\partial \tilde{r}} = \sqrt{p_{\tilde{r}}^2 + p_{\tilde{z}}^2}\,\frac{1}{2\tilde{r}^2} = \frac{H}{\tilde{r}} \qquad 13$$

$$\frac{dp_{\tilde{z}}}{d\tau} = -\frac{\partial H}{\partial \tilde{z}} = 0 \qquad 14$$

This system of equations has two invariants: equation 14 shows $p_{\tilde{z}}$ to be one; H is seen to be the other because of its τime-independence.

These can be used to relate the solution $\tilde{z} = \overline{f}(\tilde{r},\tau)$ at any τime to a given profile $\tilde{z}_i = \overline{f}_i(\tilde{r}_i,\tau_i)$ at any "initial instant" $\tau_i$. In order to achieve this, define another invariant, which can be calculated from the initial profile:

$$N \equiv \frac{p_{\tilde{z}}}{2H} = \frac{\tilde{r}}{\sqrt{p_{\tilde{r}}^2/p_{\tilde{z}}^2 + 1}} = \frac{\tilde{r}_i}{\sqrt{1 + \left(d\overline{f}_i(\tilde{r}_i,\tau_i)/d\tilde{r}_i\right)^2}} = \tilde{r}_i \cos\phi_i \equiv N_i \qquad 15$$

Note that $\phi_i$, defined in 2 and 15, is the angle made by the normal to the *initial* GV surface with the z-axis. From 12,

$$\frac{d\tilde{z}}{d\tau} = \frac{N}{2\tilde{r}^2} \qquad 16$$

From the definition of H, $p_{\tilde{r}} = \pm\sqrt{4\tilde{r}^2 H^2 - p_{\tilde{z}}^2}$, which combined with 11 gives

$$\frac{d\tilde{r}}{d\tau} = \pm\frac{1}{2\tilde{r}^2}\sqrt{\tilde{r}^2 - N^2} \qquad 17$$

From 16 and 17,

$$\frac{d\tilde{z}}{d\tilde{r}} = \frac{d\tilde{z}}{d\tau}\bigg/\frac{d\tilde{r}}{d\tau} = \pm\frac{N}{\sqrt{\tilde{r}^2 - N^2}} \qquad 18$$

For $N \neq 0$, integration of 18 yields the relation describing the family of characteristic curves, (curves of constant N and H) which are everywhere perpendicular to the GV surface (the integral surface of 7):

$$\frac{\tilde{z}}{N} + s\text{ArcCosh}\left(\frac{\tilde{r}}{|N|}\right) = C_1 = \text{Constant} \qquad 19$$

The symbol s stands for the sign of $\left(d\overline{f}_i(\tilde{r}_i,\tau_i)/d\tilde{r}_i\right)$ in its domain of definition. Also for $N \neq 0$, integration of 17 yields the location of the GV surface along the characteristic curve 19 at τime τ

$$\frac{\tilde{r}}{N}\sqrt{\frac{\tilde{r}^2}{N^2} - 1} + \text{ArcCosh}\left(\frac{\tilde{r}}{|N|}\right) + \frac{s\tau}{N^2} = C_2 = \text{Constant} \qquad 20$$

The initial profile $\tilde{z}_i = \overline{f}_i(\tilde{r}_i,\tau_i)$ provides the values of the constants $C_1$ and $C_2$ in terms of $N_i(\tilde{r}_i,\tilde{z}_i;\tau_i)$ given by 15:

$$C_1\left(\tilde{r}_i, \tilde{z}_i; N_i\right) = \frac{\tilde{z}_i}{N_i} + s\,\text{ArcCosh}\left(\frac{\tilde{r}_i}{|N_i|}\right) \qquad 21$$

$$C_2\left(\tilde{r}_i, \tau_i, N_i, s\right) = \frac{\tilde{r}_i}{N_i}\sqrt{\frac{\tilde{r}_i^2}{N_i^2} - 1} + \text{ArcCosh}\left(\frac{\tilde{r}_i}{|N_i|}\right) + \frac{s\tau_i}{N_i^2} \qquad 22$$

Equation 19 can be written in a parametric form by defining $\alpha/2 \equiv sC_1 - s\tilde{z}/N$ as

$$\tilde{r} = N\cosh(\alpha/2); \tilde{z} \equiv N(C_1 - s\alpha/2) \qquad 23$$

The GV surface is represented by the curve 23 in $(\tilde{r}, \tilde{z})$ space connecting the anode (or insulator) at radius a (or $R_I$, the insulator radius) to cathode radius $R_C$, with the value of $\alpha(\tau)$ for any $\tau$ found from the following equation derived from 22 and 20:

$$\sinh(\alpha(\tau)) + \alpha(\tau) = 2\left(C_2(\tilde{r}_i, \tau_i, N_i) - \frac{s\tau}{N_i^2}\right) \qquad 24$$

The algorithmic implementation of the above solution requires a detailed discussion, which is reserved for the next section. Only those solutions of the partial differential equation 7 which satisfy the initial and boundary conditions of the DPF can be taken to represent the GV surface.

There is a physical constraint which the GV surface needs to obey during its evolution. Assuming that all the current supplied through the anode enters the plasma at the curve of intersection between the GV surface and anode, the in-surface component of current density at the anode surface must be zero at the curve of intersection. Since the current density vector must be continuous at the anode-plasma boundary, the GV surface must meet the anode at a right angle, *whatever be the shape of the anode*.

The initial condition in DPF is usually a line in $(\tilde{r}, \tilde{z})$ space which lies over the exposed surface of the insulator: for example, the line marked ABC in Fig.1 The line has two straight sections, on which the angle $\phi_i$ has a unique value of either 0 or 90°, and three vertices A, B and C at each of which the angle $\phi_i$ takes all possible values in a continuous manner. Clearly, in order to satisfy the boundary condition, the GV surface must be constructed to be perpendicular to all the characteristic curves which can be drawn from all the points of the initial current profile. This exercise is highly non-trivial and probably represents one of the reasons why the GV model has not been used much in DPF research.

The above discussion holds true for any "initial instant" $\tau_i$. For example, at the end of the rundown phase at τime $\tau_R$, the GV surface encounters the vertex D at the end of anode. The requirement that the GV surface remain perpendicular to the anode then translates into a turn-around motion at the vertex D. This change in the character of the motion can be related to the solution described by equations 19-24 by treating the vertex D at τime $\tau_R$ as the initial GV surface and drawing characteristics from it in all possible directions.

The GV model yields simple analytic formulas in some special cases which are relevant to DPF research, which are discussed below assuming $\tau_i = 0$. More complex analytical approximations are discussed in [20].

a. <u>Lift-off from a cylindrical (insulating) surface</u>: This is the case of $\phi_i = \pi/2; N_i = 0$, which is excluded in the integrations leading to 19 and 20. The PCS shape has no dependence on the axial coordinate. Equation 8 then has the solution:

$$\psi_0 = \pm \tilde{r}^2/2 + \text{constant} \qquad \qquad 25$$

The GV surface $\psi(\tilde{r},\tilde{z};\tau) = \tau/2 + \psi_0(\tilde{r},\tilde{z}) = 0$ acquires the form

$$\tilde{r} = \sqrt{1 \pm \tau} \qquad \qquad 26$$

for the initial condition $\tilde{r} = 1$ at τ=0. The positive sign corresponds to the inverse-pinch in DPF. The GV surface reaches the cathode, of inner radius $\tilde{r}_C$ in units of anode radius, in τime $\tau_{\text{LIFTOFF}} = \tilde{r}_c^2 - 1$. The negative sign in 26 corresponds to an inward-moving snowplow sheath generated at the inner wall of a tube as in the classical snowplow z-pinch. When the initial plasma is created on the surface of an insulator with scaled radius $\tilde{r}_I$, 26 becomes $\tilde{r} = \sqrt{\tilde{r}_I^2 + \tau}$ and $\tau_{LIFTOFF} = \tilde{r}_c^2 - \tilde{r}_I^2$.

b. <u>Axial propagation in the run-down phase of coaxial plasma gun and DPF</u>: The run-down phase is defined by the physical boundary condition that at the anode radius $\tilde{r} = 1$, the plasma current sheath must have a normal intersection with the electrode surface: the case of $\phi_i = 0; N_i = 1; s = -1$. Substituting in 19 and 20, the shape-maintaining solution 10 is recovered:

$$\tilde{z} = \tilde{z}_i + \tfrac{1}{2}(\tau - \tau_i) + \tfrac{1}{2}\text{ArcCosh}(\tilde{r}) - \tfrac{1}{2}\tilde{r}\sqrt{\tilde{r}^2 - 1} \qquad 27$$

At the anode surface, the sheath takes time $\tau_R = 2(\tilde{z}_A - \tilde{z}_I)$ to reach the end of anode of length $\tilde{z}_A$ starting at the top of insulator of height $\tilde{z}_I$, both in units of anode radius.

The time $\tau$ is proportional to the charge that has flown in time t, which has to be found by solving the circuit equation. For this purpose, the inductance of the plasma can be calculated from the magnetic flux enclosed between the current sheath and the electrodes:

$$L_P = \frac{\mu_0 a}{2\pi} \iint d\tilde{z} d\tilde{r} \frac{1}{\tilde{r}} \equiv \frac{\mu_0 a}{2\pi} \mathcal{L}(\tau) \qquad 28$$

The circuit equation for a DPF driven by a capacitor bank of capacitance $C_0$, internal inductance $L_0$ and internal resistance $R_0$ charged to voltage $V_0$ is

$$\frac{d}{dt}(LI) = V_0 - \frac{1}{C_0}\int_0^t I(t')dt' - IR_0 \qquad 29$$

In terms of the 'time' $\tau$ and other dimensionless quantities: $\varepsilon \equiv Q_m/C_0 V_0$, $\kappa \equiv \mu_0 a/2\pi L_0$, $\gamma \equiv R_0\sqrt{C_0/L_0}$, $I_0 \equiv V_0\sqrt{C_0/L_0}$, $\tilde{I}(\tau) \equiv I(\tau(t))/I_0$, $\Phi \equiv (LI)/L_0 I_0 = (1 + \kappa\mathcal{L}(\tau))\tilde{I}$, this can be cast in the form

$$\Phi\frac{d\Phi}{d\tau} = \varepsilon(1 + \kappa\mathcal{L}(\tau))(1 - \varepsilon\tau) - \varepsilon\gamma\Phi \qquad 30$$

GV restrict themselves to the case of zero capacitor bank resistance ($\gamma = 0$) for which 30 gives

$$\Phi_0^2 = 2\varepsilon \int_0^\tau d\tau' \left(1 + \mathfrak{L}(\tau')\kappa\right)\left(1 - \varepsilon\tau'\right)$$
$$= 2\varepsilon\tau - \varepsilon^2\tau^2 + 2\varepsilon\kappa m_0(\tau) - 2\varepsilon\kappa\varepsilon m_1(\tau) \quad\quad 31$$
$$m_0(\tau) \equiv \int_0^\tau d\tau' \mathfrak{L}(\tau'); m_1(\tau) \equiv \int_0^\tau d\tau' \tau' \mathfrak{L}(\tau')$$

The case of non-zero circuit resistance is clearly amenable to a successive approximation scheme applied to equation 30 involving the small parameter $\varepsilon\gamma$. The flux function $\Phi(\tau)$ can be considered as the limit of a sequence of functions $\Phi_n(\tau)$, $n = 0, 1, 2 \cdots$ satisfying the equation

$$\Phi_{n+1} \frac{d\Phi_{n+1}}{d\tau} = \varepsilon\left(1 + \mathfrak{L}(\tau)\kappa\right)\left(1 - \varepsilon\tau\right) - \varepsilon\gamma\Phi_n$$
$$\Rightarrow \Phi_{n+1}^2(\tau) = \Phi_0^2(\tau) - 2\varepsilon\gamma \int_0^\tau d\tau \Phi_n \quad\quad 32$$

For values of $\varepsilon\gamma$ encountered in practice, the second iteration gives a reasonably accurate estimate of $\Phi$. The function $\mathfrak{L}(\tau)$ in the integrand of 31 is obtained explicitly from the form of the GV surface as a function of $\tau$ using the procedure already described. The connection between the dimensionless time $\tau$ and real time t, normalized to the short-circuit quarter-cycle time $T_{1/4} \equiv \pi/2 \cdot \sqrt{C_0 L_0}$, is obtained by quadrature:

$$\tilde{t} \equiv t/T_{1/4} = (2\varepsilon/\pi) \cdot \int_0^\tau d\tau'/\tilde{I}(\tau') = (2\varepsilon/\pi) \cdot \int_0^\tau d\tau' \left(1 + \kappa\mathfrak{L}(\tau)\right)/\Phi(\tau') \quad\quad 33$$

The fractions of stored energy $W_0$ converted into magnetic energy $W_m$, spent in electromagnetic work $W_W$, dissipated in circuit resistance $W_R$ and that remaining in the capacitor $W_C$ are obtained from the relations

$$\eta_m \equiv W_m/W_0 \equiv \frac{\frac{1}{2} L(t) I^2(t)}{\frac{1}{2} C_0 V_0^2} = \Phi(\tau)^2 / \left(1 + \kappa\mathfrak{L}(\tau)\right) \quad\quad 34$$

$$\eta_W \equiv W_W/W_0 \equiv \frac{1}{\tfrac{1}{2}C_0V_0^2}\int \frac{1}{2}I^2 dL = \frac{1}{\tfrac{1}{2}C_0V_0^2}\left(\int Id(LI) - \frac{1}{2}LI^2\right) \qquad 35$$
$$= \varepsilon\tau(2-\varepsilon\tau) - \eta_m - 2\varepsilon\gamma\int d\tau \tilde{I}$$

$$\eta_R \equiv W_R/W_0 \equiv \frac{R\int dt I^2(t)}{\tfrac{1}{2}C_0V_0^2} = 2\varepsilon\gamma\int d\tau \tilde{I} \qquad 36$$

$$\eta_C \equiv W_C/W_0 \equiv \frac{1}{2C_0}\left(C_0V_0 - \int_0^t I(t')dt'\right)^2 = (1-\varepsilon\tau)^2 \qquad 37$$

GV model does not take into account the delay between the start of current and the beginning of plasma lift-off. *Hence, it is impossible for the relations 31 and 33 to provide as a good match to experimentally measured current waveforms as the Lee model without introducing ad-hoc corrections*: attempts to look for quantitative agreement between the GV model and experiments without some kind of 'adjustment' are therefore *expected* to fail.

The purpose of recapitulating the GV model is to bring out the following points:

1. The only input it requires is geometrical and operational parameters and the snowplow hypothesis. *It requires no information or assumptions concerning plasma dynamics within the PCS*.

2. Its structure is entirely analytical; *there are no adjustable parameters* to bring its results closer to the experiments, except perhaps the origin of time.

3. Still, the *gross features* of the experimentally observed shape and position of the GV surface [18], of the current versus time profile [20] and inductance versus time profile [27] are reproduced.

This supports the hypothesis of decoupling between the global and local plasma dynamics [3] and *suggests the possibility that plasma phenomena within the PCS can be treated as a local flow problem in a frame-of-reference attached with the GV surface* leading to extension of the GV model. This is first of the two central themes of this paper and is briefly discussed in section IV.

The second significant aspect of the GV model is that it presents a global view of the energy transfer from the capacitor bank to the plasma using the snowplow effect in terms of seven dimensionless parameters: $\varepsilon \equiv Q_m/V_0 C_0$, $\kappa \equiv \mu_0 a/2\pi L_0$, $\gamma \equiv R_0\sqrt{C_0/L_0}$, $\tilde{z}_A, \tilde{r}_C, \tilde{r}_I$ and $\tilde{z}_I$. The last four, anode length, cathode radius, insulator radius and insulator length, all normalized to anode radius, are *independent of the scale of the device* and determine the inductance profile $\mathfrak{L}(\tau)$; this is discussed further in the next section. The first three, which involve only operational parameters and the scale of the device, determine, along with $\mathfrak{L}(\tau)$, the transformation of energy stored in the capacitor into magnetic and kinetic energy. *Trends in global maximization of energy transfer can be then analytically studied in terms of device data*; the optimum parameter range so determined can perhaps be further studied using the Lee model calibrated with a device constructed using the suggested optimum set of parameters. This aspect is discussed in more detail in section V.

### III. **Algorithmic aspects and the physical meaning of the GV model:**

This section takes a detailed look at the procedure used to construct the GV surface from given device data and its physical meaning. In its simplest manifestation, the Mather type DPF is assumed to have the anode as a straight solid cylinder of radius 1 and height $\tilde{z}_A$, the insulator is taken to be a straight cylinder of outer radius $\tilde{r}_I$ and height $\tilde{z}_I$ and cathode is taken to be a straight cylinder of inner radius $\tilde{r}_C$. The remaining device data are represented by the dimensionless numbers $\varepsilon \equiv Q_m/V_0 C_0$, $\kappa \equiv \mu_0 a/2\pi L_0$, $\gamma \equiv R_0\sqrt{C_0/L_0}$ and the discharge current is normalized to $I_0 \equiv V_0\sqrt{C_0/L_0}$.

At its simplest, the GV model treats the DPF geometry as a collection of straight lines (see Fig. 1) joined at vertices in cylindrical $(\tilde{r}, \tilde{z})$ space; the straight portions have a well-defined value of the invariant $N_i \equiv \tilde{r}_i \cos\phi_i$ while $\phi_i$ takes all possible values at the vertices. The insulator region comprises a straight portion $\tilde{r}_i = \tilde{r}_I$ for $0 \leq \tilde{z}_i \leq \tilde{z}_I$, where $N_i = 0$ and the special solution 26 applies, indicating a radially expanding cylindrical GV surface and straight-line characteristics. This surface touches the cathode at radius $\tilde{r}_C$ at $\tau_{LIFTOFF} = \tilde{r}_c^2 - \tilde{r}_I^2$; fig.1, region I shows positions of the GV surface at intervals of $\tau_R/20$. This straight portion ends in the vertex B at $(\tilde{r}_I, \tilde{z}_I)$, where $N_i$ takes values in the range $0 < N_i \le \tilde{r}_I$. Characteristics have a curved shape given by 19 and 21; the GV surface, given by 23 and 24 with s=−1 at intervals of $\tau_R/20$, is shown in Fig.1, region II. After this vertex, there is a straight line segment $\tilde{z}_i = \tilde{z}_I$ for $1 \le \tilde{r}_i \le \tilde{r}_I$ where $1 \le N_i \le \tilde{r}_I$, which ends in the vertex C at $(1, \tilde{z}_I)$. A single characteristic line emanating from this vertex is shown and the corresponding GV surface at intervals of $\tau_R/20$ is shown in fig.1, region III.

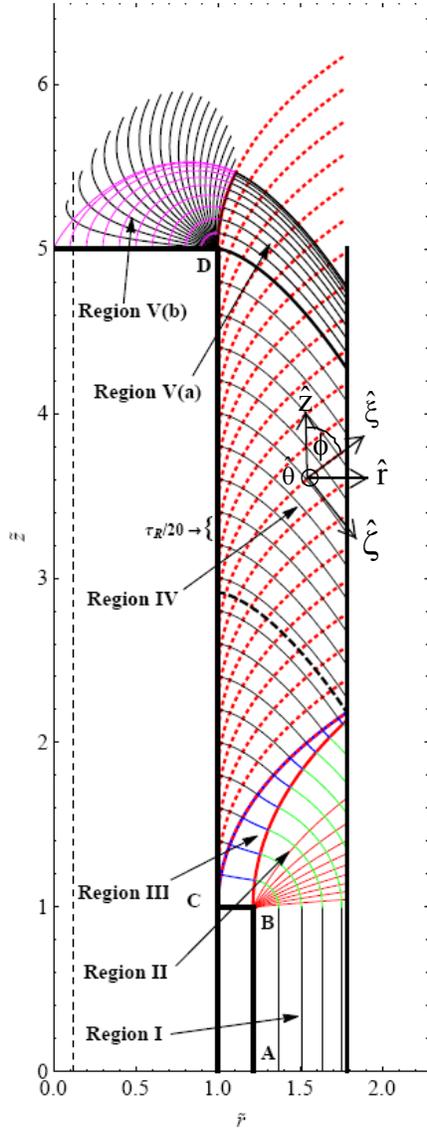

Fig. 1: Construction of characteristics and GV surface for a Mather-type plasma focus. Parameters of LPP device [12] are used for illustration: Anode radius 28 mm, anode length 140 mm, insulator radius 34 mm, insulator length 28 mm, cathode inner radius 50 mm. Unit vectors $(\hat{r}, \hat{\theta}, \hat{z})$ in cylindrical coordinate system and $(\hat{\xi}, \hat{\theta}, \hat{\zeta})$ in the local coordinate system are illustrated.

This vertex leads to the bottom of the free length of anode, over which, $\tilde{r}_i = 1, N_i = 1$. The characteristics in this region, given by 23, with $\alpha$ varying between 0 and $2\mathrm{ArcCosh}(\tilde{r}_C)$ and $\tilde{z}_i$ taking values at intervals of $(\tilde{z}_A - \tilde{z}_I)/20$ above $\tilde{z}_I$, are tangent to the anode. The GV surface given by 27 is evaluated for $\tau$ between 0 and $\tau_R = 2(\tilde{z}_A - \tilde{z}_I)$ and $\tilde{r}$ between 1 and $\tilde{r}_C$ and displayed in Fig 1, region IV.

The straight portion of anode ends in the vertex D at $(1, \tilde{z}_A)$, which is followed by the flat top portion. At the vertex D, $N_i$ varies from 1 at the straight portion to 0 on the top. The GV surface is plotted for $\tau$ given by $(\tau_R + 1) - 0.01k^2; k = 0, 1, 2 \cdots, 10$ in fig. 1, region V. The characteristic at the vertex D divides the GV surface in two regions: Region V(a) is a continuation of the separable solution of region IV, where $\tilde{r}$ varies between the characteristic at the vertex and $\tilde{r}_C$ and region V(b) is described by variation of $N_i$ between 0 and 1. The GV surface at $\tau = \tau_R + 1$ passes through $\tilde{r} = 0$; it does not predict a pinch column of finite radius. This is expected because the snowplow model concerns a moving region while the pinch column is a static phenomenon. Empirical data [1] shows that the pinch radius is given by $\tilde{r}_p \approx 0.12$ (shown by vertical dashed line in fig.1); this value is reached at $\tau_P \approx \tau_R + 0.986$, beyond which, the GV model loses its physical context. The shape of GV surface near $\tau_P$ has a noteworthy resemblance with plasma shape observed in schlieren pictures of DPF [28]. Many times, the anode is a hollow tube with one more vertex, leading to a turn-around motion into the tube. This interesting case is discussed in [20] but is omitted for lack of space and relevance.

The GV model is seen to provide a relation between the device geometry and the region accessible to the snowplow motion *and hence its inductance*. For given device geometry data, $\mathcal{L}(\tau)$ can be determined numerically and fitted to simple forms; the fitting parameters can themselves be fitted as functions of the aspect ratios $\tilde{z}_A, \tilde{r}_C, \tilde{r}_I$ and $\tilde{z}_I$. GV have given such fitted formulas in [20]; it is apparent however that their fitted forms refer to some unspecified sets of aspect ratios since their formula predicts the inductance contribution from Region II to be independent of the normalized cathode radius, which is impossible.

Parameter space studies using automated computations, few thousands in number, over the range $1.01 \leq \tilde{r}_I \leq 1.04$, $0.5 \leq \tilde{z}_I \leq 2$, $\tilde{r}_I + 0.2 \leq \tilde{r}_C \leq 2.0$, $\text{Max}\left[2, \tilde{r}_C, 0.5\tilde{r}_C^2 - 1 + \tilde{r}_I\right] \leq \tilde{z}_A \leq 10 + \tilde{r}_I$, which may be thought to represent the commonly accepted conception of "Mather type" DPF, show that $\mathcal{L}(\tau)$ can be represented by the following expression similar to the GV expression for inductance

$$\mathcal{L}(\tau) = \tilde{z}_I \text{Log}\left(\sqrt{\tilde{r}_I^2 + \tau}\right) + k_1 \text{Log}(\tilde{r}_C)\tau^{1.5} \quad 0 < \tau \leq \tau_{\text{LIFTOFF}}$$

$$= \mathcal{L}(\tau_{\text{LIFTOFF}}) + \frac{1}{2}(\tau - \tau_{\text{LIFTOFF}})\text{Log}(\tilde{r}_C) + k_2 \text{Log}(\tilde{r}_C) \quad \tau_{\text{LIFTOFF}} \leq \tau \leq \tau_R \qquad 38$$

$$= \mathcal{L}(\tau_R) - k_3 \text{Log}(\tau_R + 1 - \tau) \quad \tau_R \leq \tau \leq \tau_R + 0.98$$

The three parameters $k_1$, $k_2$, $k_3$ are found to be independent of the anode length $\tilde{z}_A$, insulator length $\tilde{z}_I$ and insulator radius $\tilde{r}_I$ all scaled to anode radius. They depend on the scaled cathode radius $\tilde{r}_C$ as

$$k_1 = \frac{\lambda_0}{\tilde{r}_c + \lambda_1}; k_2 = \lambda_2 + \lambda_3 \tilde{r}_c + \lambda_4 \tilde{r}_c^2; k_3 = \lambda_5 + \lambda_6 \tilde{r}_c + \lambda_7 \tilde{r}_c^2 \qquad 39$$

$\lambda_0$=0.276304; $\lambda_1$=−0.68924; $\lambda_2$=−0.08367; $\lambda_3$=0.105717; $\lambda_4$=−0.02786;

$\lambda_5$=−0.05657; $\lambda_6$=0.263374; $\lambda_7$=−0.04005;

Formulas 38 and 39 represent the parametric summary of GV model for the Mather-type DPF.

Fig. 2 shows an example of inductance variation calculated from the code and compared with the above modification of GV's fitted formula. It resembles experimental inductance profiles [27]. Fig. 3 shows comparison of current profile calculated from the resistive extension of the GV model with published experimental data.

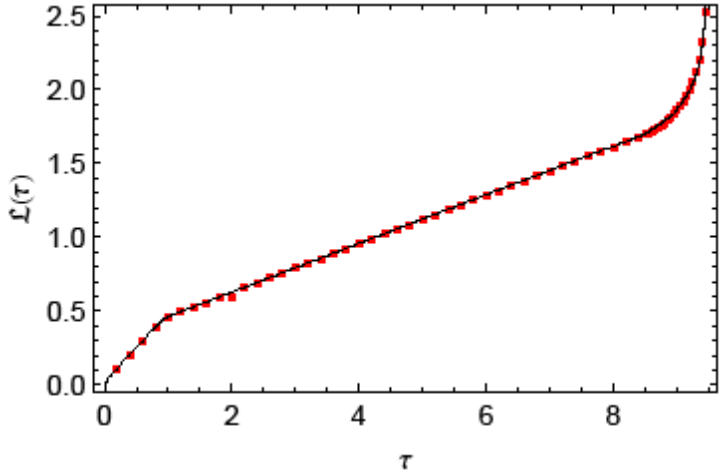

Fig.2: Inductance variation in GV model (points) and its fit to 38 (solid line). The fitted parameters are: $k_1$=0.393559, $k_2$=0.00948515, $k_3$=0.232337. Device parameters correspond to PF-1000 (see fig 3): $\tau_{\text{LIFTOFF}} \approx 0.918$, $\tau_R = 8.47$.

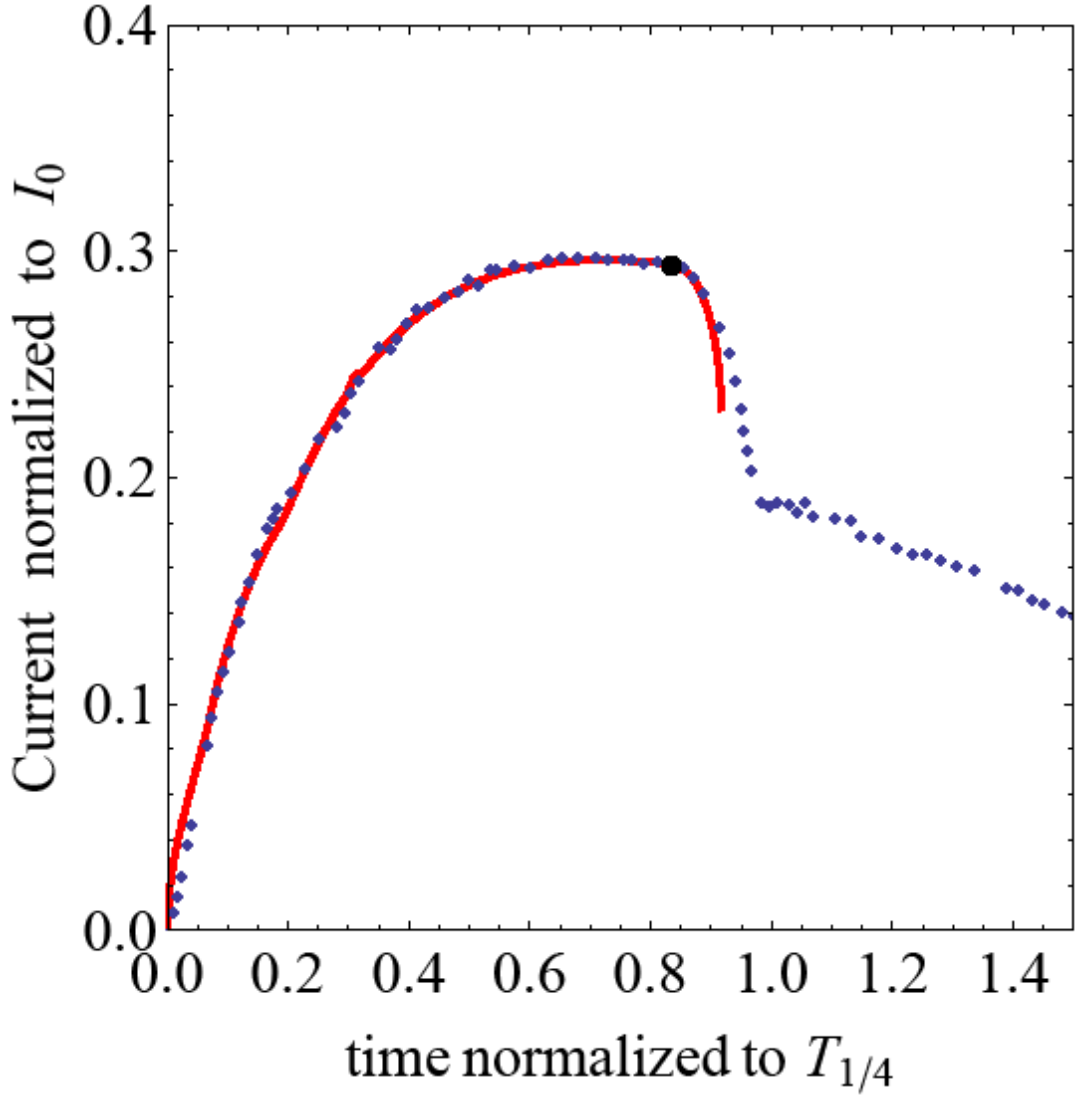

Fig. 3: Comparison of current profile from resistive extension of GV model (solid line) with experimental data (points) of Gribkov et. al. [29], digitized and used for comparison with the Lee model in [30]. The parameters of fig. 3(a) are: $C_0$=1332 µF, $V_0$=27 kV, anode radius a= 115 mm, anode length =600 mm, insulator radius=116 mm, insulator length=113 mm, cathode inner radius =160 mm. Three parameters, not mentioned in [29], were varied to get a good fit: $L_0$=25 nH, $D_2$ pressure=2.6 torr, circuit resistance =5.5 mΩ; the origin of time was *not* altered. Note that the Lee model fit [30] occurs at different values of parameters and covers a larger portion of the current profile. Scholz et. al. [31] report comparison of experimental current waveform of PF-1000 with several models and the closest model is snowplow model with inductance 20 nH and resistance 5.5 mΩ. End-of-rundown is indicated by the dot on the curve; at this time $\eta_M = 0.22$, $\eta_W = 0.10$, $\eta_C = 0.48$, $\eta_R = 0.20$. At the end-point of the GV model, $\eta_M = 0.19$, $\eta_W = 0.15$, $\eta_C = 0.43$, $\eta_R = 0.23$.

Another important aspect of the GV model is the volume swept by the PCS during its motion. Energy spent in ionizing and displacing the gas contained in this volume must be supplied through the electromechanical work represented by 35. The volume swept up to $\tau_R$ is given in the notation of this paper by the formula [20],

$$\upsilon = \pi a^3 \left(\tilde{r}_c^2 - 1\right)\left(\tilde{z}_A - \tfrac{3}{7}h\right); \quad h = \tfrac{1}{2}\tilde{r}_c\sqrt{\tilde{r}_c^2 - 1} - \text{Log}\left(\tilde{r}_c + \sqrt{\tilde{r}_c^2 - 1}\right) \qquad 40$$

It is clear that the GV formalism describes the mathematical consequences of the snowplow hypothesis in the context of a certain construction of electrodes: *it is inherently not a model of any plasma*. As a mathematical theory, it shows the relation between the device geometry, the region accessible to the snowplow plasma (and hence its inductance) and the time required for the evolution. As shown in [20], it provides an algorithm for handling somewhat more complex evolution involving inward motion into a hollow anode and formation of an expanding bubble. It is a powerful mathematical structure without any adjustable parameters and *without any physical content beyond that given by 1*. The resemblance between some predictions of this model, such as the shape of the GV surface and current profile and corresponding observations on DPF merely tend to suggest that in *some operational regimes, DPF sheath evolution is well-approximated by the snowplow model*. It is then reasonable to explore whether this model can be extended to reproduce qualitative features of the sheath structure in the same operational regimes; this forms the focus of the next section. Conversely, when the GV model predictions do not match experimental data, it probably means that phenomena other than the snowplow effect are playing an important role; *the discrepancy may then serve as a diagnostic tool [22]*.

### IV. Extension of the GV model: sheath structure modeled as a local plasma flow

This section introduces the basic ideas concerning the possibility of extension of the GV model; no attempt is made *in this paper* to explore the extended formalism. The purpose is to demonstrate the significance of the *analytical nature* of the GV model as opposed to the purely numerical nature of other models.

The GV model may be extended by incorporating additional simplifying assumptions.

1. The relative velocity between the front and back surfaces of the PCS, representing boundaries between current-carrying and current-free zones, may be neglected in comparison with the velocity of the GV surface, so that they may be considered "rigidly attached to each other" in a first approximation. Both the surfaces should then belong to the family of curves in (r,z) space orthogonal to the characteristics given by 19. In this picture, the balance between "wind pressure" at the front surface $\Sigma_1$ and magnetic pressure at the back surface $\Sigma_2$ represents a boundary condition on the plasma dynamics within the 'rigid' region bounded by these two surfaces and the electrodes. *One consequence of this boundary condition is that the component of plasma velocity along the characteristics between $\Sigma_1$ and $\Sigma_2$ must be equal to the normal velocity of the GV surface.*

2. The density of plasma behind the PCS is known to be much less than the density of the fill gas and the density of the plasma ahead of the PCS is known to be of constant order of magnitude during the rundown and radial collapse phase. Mass flowing along the normal into the PCS must therefore exit the PCS *tangentially* so that it neither accumulates within it nor is admitted into the region behind it.

3. The pressure gradient and magnetic force are assumed to act mainly along the normal to the PCS and forces acting in the tangential direction are neglected to first order and are left to be determined as second order corrections by successive approximation. This assumption is equivalent to the assertion that the tangential flow is driven primarily by the curvature of the PCS, *which is captured in the analytical structure of the GV model*.

4. The local plasma dynamics is assumed to be much faster than the timescale of the PCS evolution so that it may be assumed to have reached a quasi-stationary state: its time evolution should be mainly governed by the motion of the PCS.

5. Axial magnetic field, azimuthal current density and azimuthal plasma velocity are assumed to be second order effects, which can be neglected initially and determined as corrections by successive approximation.

These simplifying assumptions allow formulation of a first order model of sheath structure by constructing a local coordinate system $(\xi,\theta,\zeta)$ (see Fig. 1) attached with the GV surface in terms of a unit vector along the normal defined as $\hat{\xi} \equiv \vec{\nabla}\psi / |\vec{\nabla}\psi| = \hat{r}\tilde{r}^{-1}\sqrt{\tilde{r}^2 - N^2} + \hat{z}\tilde{r}^{-1}sN$, a unit vector $\hat{\theta}$ along the azimuth and a unit vector along the tangent to GV surface defined as $\hat{\zeta} \equiv \hat{\xi} \times \hat{\theta} = \hat{z}\tilde{r}^{-1}\sqrt{\tilde{r}^2 - N^2} - \hat{r}\tilde{r}^{-1}sN$. In this coordinate system, differential of the coordinate vector is represented by:

$$d\vec{r} = \hat{r}dr + \hat{\theta}rd\theta + \hat{z}dz = \hat{\xi}d\xi + \hat{\theta}(r(\xi,\zeta)d\theta) + \hat{\zeta}d\zeta \qquad 41$$

Substituting the expressions for $\hat{\xi}$ and $\hat{\zeta}$ and equating $\hat{r}$, $\hat{z}$ components one gets

$$d\xi = \tilde{r}^{-1}\sqrt{\tilde{r}^2 - N^2}d\tilde{r} + \tilde{r}^{-1}sNd\tilde{z}$$
$$d\zeta = -\tilde{r}^{-1}Nsd\tilde{r} + \tilde{r}^{-1}\sqrt{\tilde{r}^2 - N^2}d\tilde{z} \qquad 42$$

Along the normal, $d\zeta = 0$, establishing a relation between $d\tilde{r}$ and $d\tilde{z}$, which gives

$$d\xi = \frac{\tilde{r}d\tilde{r}}{\sqrt{\tilde{r}^2 - N^2}} \qquad 43$$

Similarly, along the tangent $d\xi = 0$, which gives

$$d\zeta = -\frac{s\tilde{r}}{N}d\tilde{r} \qquad 44$$

so that

$$\xi = \pm\sqrt{\tilde{r}^2 - N^2}\,;\ \zeta = -s\tilde{r}^2/2N. \qquad 45$$

Inverting, one gets

$$\tilde{r}(\xi,\zeta) = \sqrt{2\zeta^2 \pm 2\zeta\sqrt{\zeta^2 - \xi^2}}$$
$$N(\xi,\zeta) = -s\left(\zeta \pm \sqrt{\zeta^2 - \xi^2}\right) \qquad 46$$

Expressions for some vector differential operators in this system are given in the Appendix. GV scaling allows construction of dimensionless quantities:

$$\tilde{\rho} = \rho/\rho_0,\quad \tilde{v} = \vec{v}/v_A\,;\quad v_A = \mu_0 I(t)/\pi a\sqrt{2\mu_0\rho_0}\,;\quad \tilde{p} = p/p_0\,;\quad p_0 \equiv \left(\mu_0 I^2/2\pi^2 a^2\right);\quad \tilde{J} = \vec{J}/J_0\,;$$

$$J_0 \equiv \left(I/\pi a^2\right);\ \tilde{B} = B(\tilde{r},\tilde{z})\hat{\theta}/B_0\,,\ B_0 = \mu_0 I/2\pi a\,;\ \tilde{\nabla} = a^{-1}\vec{\nabla}\,;\ \Lambda = 2m_i m_e/\rho_0\mu_0 e^2$$

The equations of continuity and momentum conservation then become [32]:

$$\tilde{\nabla} \cdot (\tilde{\rho}\tilde{v}) = 0 \qquad 47$$

$$\tilde{\rho}(\tilde{v} \cdot \tilde{\nabla})\tilde{v} = \tilde{J} \times \tilde{B} - \tilde{\nabla}\tilde{p} - \Lambda\tilde{\nabla} \cdot \left(\frac{\tilde{J}\tilde{J}}{\tilde{\rho}}\right) \qquad 48$$

Note that the explicit τime dependence is dropped because of the assumption concerning the local plasma dynamics having reached a stationary state on the time scale of the PCS motion. The boundary condition referred above becomes

$$\tilde{v}_\xi = (2\tilde{r}(\xi,\zeta))^{-1} \qquad 49$$

Using assumption (3), 48 gives for the tangential component of plasma velocity

$$(\tilde{v} \cdot \tilde{\nabla})\tilde{v}\big|_\zeta = 0 \qquad 50$$

Using the expression for convective acceleration from the Appendix, 50 can be written as

$$\tilde{v}_\xi \partial_\xi \tilde{v}_\zeta + \tilde{v}_\zeta \partial_\zeta \tilde{v}_\zeta + \tilde{v}_\xi \tilde{v}_\xi \mathbb{C}(\xi,\zeta) + \tilde{v}_\zeta \tilde{v}_\xi (\mathbb{D}(\xi,\zeta) - \mathbb{C}(\xi,\zeta)) = 0 \qquad 51$$

Note that 47 can be written as

$$(\tilde{v}_\xi \partial_\xi + \tilde{v}_\zeta \partial_\zeta)\log(\tilde{\rho}) = -(\tilde{r}(\xi,\zeta))^{-1} \partial_\zeta (\tilde{r}(\xi,\zeta)\tilde{v}_\zeta) \qquad 52$$

Solution of 51, along with 49, enables solution of 52, giving the density as a function of coordinates, *without requiring any information about temperature or current profile or any processes involved in local energy conservation*. This happens because *the GV model analytically captures the curved shape of the PCS* and the tangential flow field is primarily determined by the sheath curvature in the absence of forces acting in the tangential direction. The density profile is related to the electrode geometry by the boundary conditions which require the GV surface to be normal to the anode at the curve of intersection.

The tangential flow would be very small in the neutral gas, increasing gradually as the front surface is approached and must be strong enough to eject all the mass flowing across the front surface; in other words, it would be an increasing function along $-\hat{\xi}$ direction. The logarithmic dependence on density in 52 then indicates a sharp decrease in density in the upstream region indicating an efficient snowplow sweeping action. The tangential plasma flow should play a role in preserving the stability of the curved plasma shape by 'washing away' perturbations on the same time scale as the plasma motion.

In a similar fashion, the Generalized Ohm's Law with temperature dependent resistivity, Maxwell's equations and the equation for thermal transport with Ohmic heating can be written in the local coordinate system, determining the (quasi-stationary) distributions of temperature (hence pressure), current density and magnetic field. Since the left hand side of normal component of 48 is determined by the boundary condition in terms of 49 and the solution of 51, the normal component of equation 48 can be balanced only if one introduces the concept of free currents [4], which act as sources of axial magnetic field in second order of successive approximation. This potentially leads to a theory of the *azimuthally symmetric* sheath structure consistent with observation of an axial magnetic flux in the rundown region [4] and with solenoidal features of plasma referred earlier.

Significance of the GV scaling comes into play here: since the magnetic field has amplitude directly proportional to the current and inversely proportional to the size of the device, the vector potential must have amplitude *proportional to the current and independent of the size of the device*. At the PCS front surface, the plasma and the magnetic field are created together; clearly, canonical momentum must remain unchanged and equal to zero during this process. The canonical momentum of some electrons and ions should then be equal in magnitude to the canonical momentum of the magnetic field; their energy would then be proportional to the square of current and *independent of the dimensions of the device*; experimental evidence [33] exists supporting dependence of ion energy on square of current. The 3-dimensional geometry of vector potential, whose azimuthal component comes from the axial magnetic field, supports completely confined ion trajectories, described by constant energy and canonical momentum, of size comparable to the plasma size which scales with the dimensions of the device; this is consistent with experimental data [4] suggesting toroidal ion trajectories encircling the device axis.

This brief outline of formulation of a model of sheath structure is intended as a *demonstration of the importance of the analytical formulation of the shape of the PCS in the GV model*; details of the model are beyond the scope of this paper.

## V. Nature of the device optimization problem

The purpose of formulating the device optimization problem needs to be clearly stated and understood. At the present time, understanding of physical processes responsible for desirable outcomes of snowplow devices like DPF is inadequate. In spite of such poor understanding, the device needs to be physically realized with a set of device parameter values which makes best use of available stored energy. It is not possible to confidently predict the optimum parameters using any theoretical model, including sophisticated 3-D numerical codes, *because the underlying physics has not yet been fully established [4]*. The imperfect model, whether numerical, with its adjustable parameters or analytical, with its idealized assumptions, can at most provide initial values of parameters for an *iterative empirical optimization procedure*. The device itself needs to be engineered to provide sufficient flexibility for empirical optimization.

The snowplow effect can be looked upon as a means for delivering energy stored in the capacitor bank to the plasma *without inquiring into what the plasma does with that energy*. The GV model then allows formulation of a variety of optimization problems for an arbitrary snowplow device. It describes the energy transfer process by mapping the ten device parameters: capacitance, inductance, resistance, voltage, pressure (or density), anode radius, anode length, insulator radius, insulator length and cathode radius, on to seven dimensionless parameters $\varepsilon \equiv \pi a^2 \sqrt{2\mu_0 \rho_0} / \mu_0 V_0 C_0$, $\kappa \equiv \mu_0 a / 2\pi L_0$, $\gamma \equiv R_0 \sqrt{C_0 / L_0}$, $\tilde{z}_A, \tilde{r}_C, \tilde{r}_I$ and $\tilde{z}_I$. If the empirical optimization procedure *happens* to reveal some relationship between these dimensionless parameters, that would be a result of, and a clue to, *physics external to the GV model*. The empirically observed near-constancy of the drive parameter $I/a\sqrt{p}$ [1] for well-optimized Mather type DPF devices over 8 decades of stored energy can be cited as an example. Since the energy required to ionize the volume of gas swept by the PCS must be supplied [20,23,24,25] by the electromechanical work represented by 35, there should exist an upper pressure limit for the operation of a Mather type DPF device; this conjecture, supported by experimental data [23], is another example of physics external to the snowplow effect.

The optimization problem can be generically stated as follows: given a capacitor bank, how should the snowplow device parameters be chosen to yield the best results? This is essentially a quest for a set of constraints which should ideally be satisfied only by a unique set of device parameter values. Implicit in this statement is the idea that "best" must be defined in terms of the desired application. For example, if the end result is material modification by impact of energetic plasma, the total plasma energy (magnetic energy coupled with *plasma* inductance + work done) may need to be maximized. If the goal is to maximize fusion reaction yield in the plasma focus, the ion energy, the plasma density and the size of the plasma must be maximized together.

The optimization problem can be formulated for a zero resistance case for a chosen terminal point $\tau^*$ of the GV model, since the resistance effect can be accounted for in empirical optimization. One desirable optimization goal would be maximum conversion of energy into magnetic energy associated with plasma inductance *in minimum time* in order to minimize various energy losses from dissipative processes (such as radiation, heat conduction to electrodes, Ohmic heating in circuit resistance) neglected in the GV model. The fraction of energy converted into magnetic energy coupled with plasma inductance after τime $\tau^*$ is given by

$$\eta_{MP}(\tau^*) = \kappa \mathcal{L}(\tau) \tilde{I}^2(\tau^*)$$

Taking the example of a Mather type DPF, one may take the end of run-down phase [34] as the termination point of GV model. Fig. 4 compares the distribution of the optimization target functional $\eta_{MP}(\tau^*)/\tilde{t}(\tau^*)$, which may be called "average power parameter", in $(\varepsilon, \kappa)$ parameter space for two contemporary mega-ampere facilities: PF-1000 and LPP.

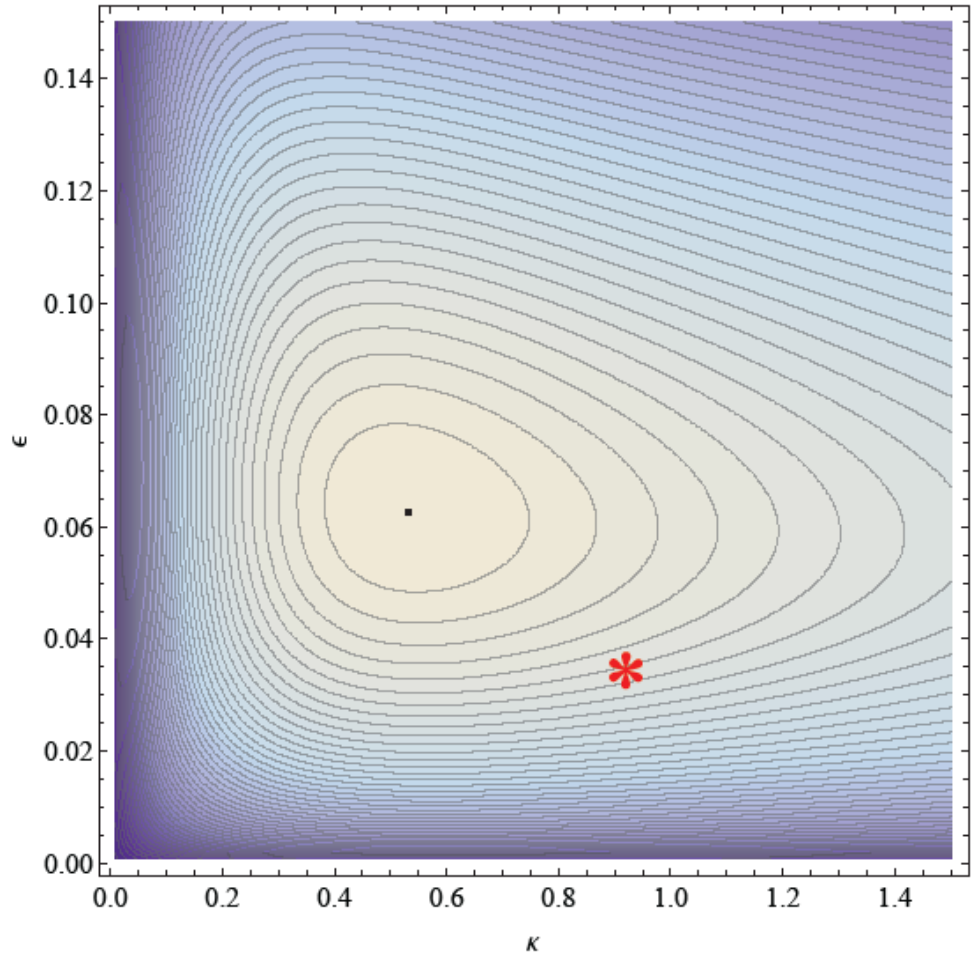

Fig 4a

Fig. 4b.

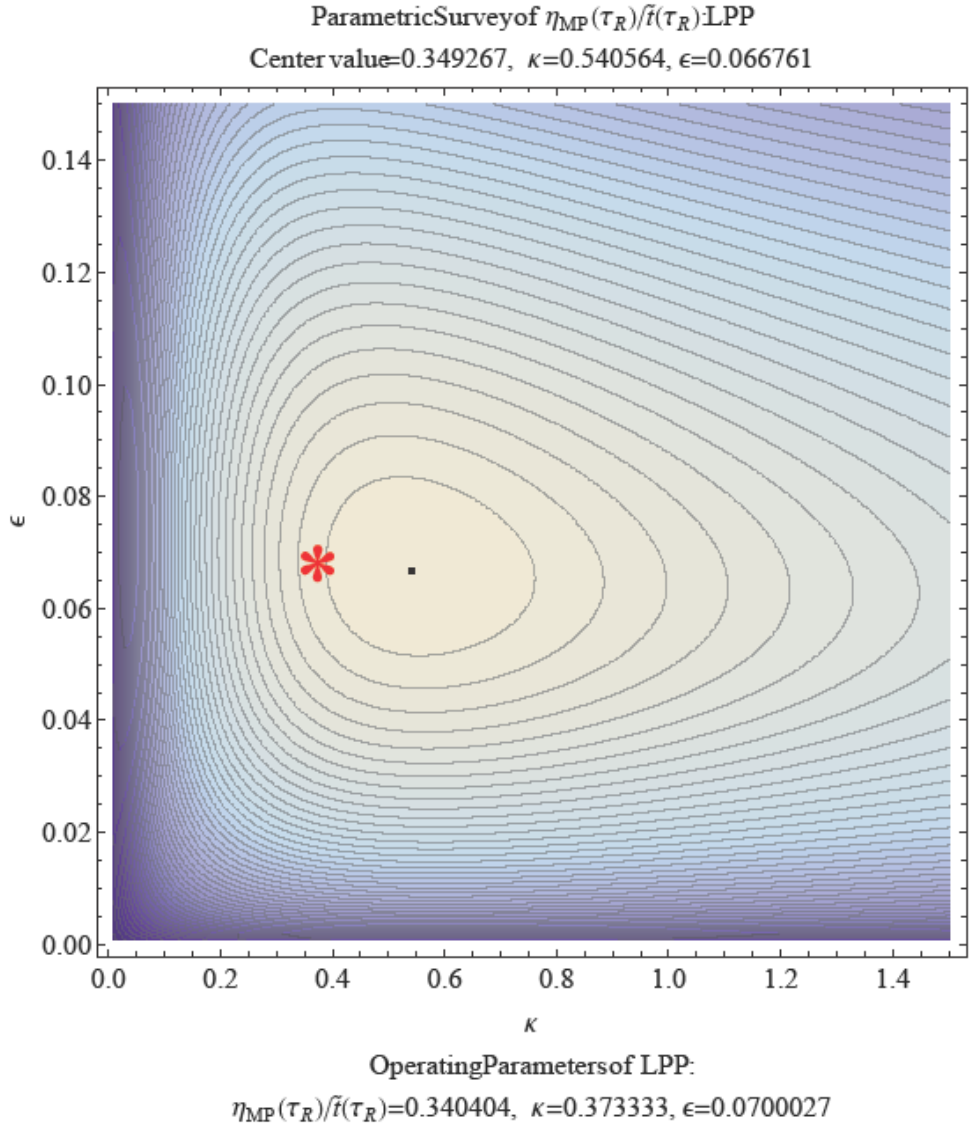

Fig 4: Comparison of distribution of the average power parameter $\eta_{MP}(\tau^*)/\tilde{t}(\tau^*)$ over 22,500 points in $(\varepsilon,\kappa)$ parameter space of zero-resistance GV model for two contemporary mega-ampere facilities: PF-1000 (parameters as in Fig 3) and LPP (pressure 24 torr, dimensions as in [12]). The operating points of these facilities are marked with an asterisk. The contours are spaced at 2% of maximum value of the average power parameter. Note that the peak values of the average power parameter differ by only 0.7%, the $(\kappa,\varepsilon)$ coordinates of the peak differ by 1.7% and 6.5%

respectively while the scaled anode length differs by 36% and scaled cathode radius differs by 18%. The zero-resistance GV model reveals the following partitioning of energy at the end of the rundown phase:

PF-1000: $\eta_M = 0.34$, $\eta_W = 0.18$, $\eta_C = 0.48$; $\eta_{MP} = 0.25$.
LPP: $\eta_M = 0.61$, $\eta_W = 0.20$, $\eta_C = 0.19$; $\eta_{MP} = 0.31$.

It is noteworthy that LPP reports [12] neutron yield significantly above the global scaling law. If one tentatively correlates these reports with the choice of an operating point closer to the optimum in $(\varepsilon, \kappa)$ space, the logical next question would be to inquire whether a global maximum of the average power parameter, or a similar quantity, exists in the parameter space consisting of all the dimensionless parameters; this question is not pursued here because of space constraints. However, two related points can be immediately addressed: firstly, existence of any such global optimum within the framework of GV model cannot explain the empirical constancy of the drive parameter *because no combination of dimensionless constants of GV model can produce a quantity which is proportional to the drive parameter and contains no other device property.* Secondly, an infinity of facilities can in principle be constructed operating at any specified point in the parameter space of the GV model. This is because of the mapping of 10 device parameters onto 7 dimensionless GV model parameters, leaving 3 extra degrees of freedom associated with every point in the 7-dimensional parameter space of the GV model. These three degrees of freedom could be chosen as capacitance, inductance and voltage of the bank; then it is possible to assert that for every capacitor bank, it is possible to construct a DPF device having a one-to-one mapping between its operational parameters and dimensionless parameters of the GV model. Conversely, there exist similarity classes of DPF devices, which are represented by the same set of dimensionless parameters in the GV model.

### VI. Summary and conclusions

Both numerical models with ad-hoc adjustable parameters and analytical models with idealized simplifying assumptions are complementary approaches to understanding the undiscovered physics of the Dense Plasma Focus device. This paper attempts to present a re-appraisal of the Gratton-Vargas (GV) analytical snowplow model of the plasma focus developed in 1970's. It is argued that the GV model is a powerful mathematical theory of the snowplow effect in a system of coaxial electrodes, with a rich, not-fully-explored structure; however, it is not a theory of any plasma and has no physical content beyond the snowplow effect. By itself, it cannot hope to reproduce any quantitative measurements of plasma properties without introducing some adjustable parameters; however it does produce plasma and current profiles which resemble those experimentally observed if the model parameters are suitably chosen.

In spite of this limitation, its transparent assumptions and analytical representation of plasma profile make the GV model a potentially useful platform for constructing more sophisticated models, whose predictions could be compared with experimental measurements and the discrepancy could be utilized to decipher the unknown physics of the Dense Plasma Focus device. As an illustration, a curvilinear coordinate system based on the parametric representation of plasma shape in the GV model has been constructed and an equation for the tangential flow field driven primarily by plasma curvature is derived. It is shown that a first order theory of plasma density structure can be constructed without requiring any information about energy transport processes or current density distribution. The approach is compatible with construction of a theory of generation of axial magnetic flux in azimuthally symmetric plasma in the rundown phase consistent with observation of an axial magnetic flux in the rundown region [4] and with solenoidal features of plasma inferred from fusion product diagnostics [4].

An important aspect of the GV model is the scaling of the device parameters. The ten device parameters: capacitance, inductance, resistance, voltage, pressure (or density), anode radius, anode length, insulator radius, insulator length and cathode radius, are mapped on to only seven dimensionless parameters $\varepsilon \equiv \pi a^2 \sqrt{2\mu_0 \rho_0} / \mu_0 V_0 C_0$, $\kappa \equiv \mu_0 a / 2\pi L_0$, $\gamma \equiv R_0 \sqrt{C_0/L_0}$, $\tilde{z}_A, \tilde{r}_C, \tilde{r}_I$ and $\tilde{z}_I$. It is because of this that no optimization problem based only upon the GV model can reproduce the observed near-constancy of the drive parameter [1]; an additional set of dimensionless parameters from physics external to the GV model would be required to form a combination proportional to the drive parameter containing no other device property. This also suggests existence of similarity classes of DPF devices: to the extent device operation can be approximated by the snowplow hypothesis, a plurality of device parameters should produce identical values of all GV model dimensionless parameters. They should then have identical current profiles when expressed in scaled units as in Fig. 3 and also similar partition of stored energy into various forms. It should then be possible to conceive of a "best possible design" in terms of dimensionless parameters for a specific application and use that to produce scaled versions at different levels of energy, physical size and cost. This has significant implications for the realization of DPF as a technology platform for commercial applications.

Towards this goal, the GV model also provides a parametric representation of dynamic inductance as a function of dimensionless parameters. A wide variety of optimization problems tailored to specific quantitative performance goals can then be formulated for snowplow devices of arbitrary axially symmetric geometry. Because of this, once the physical mechanism responsible for the significantly-higher-than-thermonuclear fusion reaction rate in DPF is understood, the GV model can help in realizing that mechanism in a technologically different and superior device.

Therefore a strong case exists for incorporation of the GV model in contemporary research programs dealing with the Dense Plasma Focus and its applications.

References

1. Leopoldo Soto, Cristian Pavez, Ariel Tarifeno, José Moreno, and Felipe Veloso, "Studies on scalability and scaling laws for the plasma focus: similarities and differences in devices from 1MJ to 0.1 J", Plasma Sources Sci. Technol. 19 (2010) 055017 (9pp).

2. S. Lee , S. H. Saw , A. E. Abdou , H. Torreblanca, "Characterizing Plasma Focus Devices—Role of the Static Inductance—Instability Phase Fitted by Anomalous Resistances" J Fusion Energ (2011) 30:277–282.

3. S.K.H. Auluck, "Description of plasma focus current sheath as the Turner relaxed state of a Hall magnetofluid" Phys. Plasmas, 16, 122505, (2009).

4. S.K.H. Auluck, "Manifestation of Constrained Dynamics in a Low-Pressure Spark", IEEE Trans. Plasma Sci., Vol. 41, No. 3, March 2013 pp. 437-446.

5. V. I. Krauz, K. N. Mitrofanov, M. Scholz, M. Paduch, P. Kubes, L. Karpinski, and E. Zielinska "Experimental evidence of existence of the axial magnetic field in a plasma focus", Eur Phy. Lett, vol 98 (2012) 45001,

6. P Kubes, V Krauz, K Mitrofanov, M Paduch, M Scholz, T Piszarzcyk, T. Chodukowski, Z Kalinowska, L Karpinski, D Klir, J Kortanek, E Zielinska, J Kravarik and K Rezac, "Correlation of magnetic probe and neutron signals with interferometry figures on the plasma focus discharge"Plasma Phys. Control. Fusion 54 (2012) 105023.

7. P Kubes, D Klir, J Kravarik, K Rezac, J Kortanek, V Krauz, K Mitrofanov, M Paduch, M Scholz, T Pisarczyk, T Chodukowski, Z Kalinowska, L Karpinski and E Zielinska, "Scenario of pinch evolution in a plasma focus discharge", Plasma Phys. Control. Fusion 55 (2013) 035011 (8pp).

8. M. Krishnan, "The Dense Plasma Focus: A Versatile Dense Pinch for Diverse Applications" IEEE TRANS. PLASMA SCI., VOL. 40, (2012) p. 3189.

9. R.S Rawat, "High-Energy-Density Pinch Plasma: A Unique Nonconventional Tool for Plasma Nanotechnology", IEEE TRANS. ON PLASMA SCIENCE, VOL. 41, NO. 4, APRIL 2013, p.701.

10. V. A. Gribkov, "Dense Plasma Focus: Current and Perspective Applications", School and Training Course on Dense Magnetized Plasma as a Source of Ionizing Radiations, their diagnostics and Applications, Abdus Salaam International Center for Theoretical Physics, Trieste, Italy, 8 - 12 October 2012.

11. G Giannini, V Gribkov, F Longo, M Ramos Aruca and C Tuniz, "Opportunities afforded by the intense nanosecond neutron pulses from a plasma focus source for neutron capture therapy and the preliminary simulation results, Phys. Scr. 86 (2012) 055801 (9pp).

12. Eric J. Lerner , S. Krupakar Murali , A. Haboub, "Theory and Experimental Program for p-$^{11}$B Fusion with the Dense Plasma Focus", J Fusion Energy 30 (2011)367–376

13. Ahmad Talaei , S.M.Sadat Kiai, A.A.Zaeem, "Effects of admixture gas on the production of 18F radioisotope in plasma focus devices", Applied Radiation and Isotopes 68 (2010) 2218–2222.

14. G. Bockle, J. Ehrhardt, P. Kirchesch, N. Wenzel., R. Batzner, H. Hinsch and K. Hübner, "Spatially resolved light scattering diagnostic on plasma focus devices", Plasma Phys. Cont Fusion, 34, pp 801-841, 1992.

15. A. Schmidt, V. Tang, and D. Welch, "Fully Kinetic Simulations of Dense Plasma Focus Z-Pinch Devices" PRL 109, 205003 (2012);

16. S. F. Garanin and V. I. Mamyshev, "Two-Dimensional MHD Simulations of a Plasma Focus with Allowance for the Acceleration Mechanism for Neutron Generation" Plasma Physics Reports, Vol. 34, No. 8, pp. 639–649. (2008).

17. V. V. Vikhrev and V. D. Korolev, "Neutron Generation from Z-Pinches", Plasma Physics Reports, vol. 33, pp. 356–380, 2007.

18. M. G. Haines, "A review of the dense Z-pinch" Plasma Phys. Control. Fusion 53 (2011) 093001 (168pp).

19. Eric J. Lerner, S. Krupakar Murali, Derek Shannon, Aaron M. Blake, and Fred Van Roessel, "Fusion reactions from >150 keV ions in a dense plasma focus plasmoid", PHYSICS OF PLASMAS 19, 032704 (2012).

20. F. Gratton and J.M. Vargas, "Two dimensional electromechanical model of the plasma focus", in Energy Storage, Compression and Switching, V. Nardi, H. Sahlin, and W. H. Bostick, Eds., vol. 2. New York: Plenum, 1983, p. 353.

21. J. M. Vargas, F. Gratton, J. Gratton, H. Bruzzone, and H. Kelly, "Experimental verification of a theory of the current sheath in the plasma focus." in Proc. 6th Int. Conf. on Plasma Phys. and Controlled Fusion Res., Berchtesgaden, 1976, IAEA-CN-35/E18.5(a), p. 483.

22. H Bruzzone and J F Martınez, "Cinematic of the current sheet in a pulsed coaxial plasma source operated with uniform gas filling", Plasma Sources Sci. Technol. 10 (2001) 471–477.

23. M. Milanese, J. Pouzo, "Critical Analysis of Plasma Focus Design Based on the Implications of an Upper Pressure Limit", Nuclear Fusion, 25, 7, pp. 840 - 846 (1985).

24. J. Pouzo, D. Cortázar, M. Milanese, R. Moroso, R. Piriz, "Limits of Deuterium Pressure Range with Neutron Production in Plasma Focus Devices", Small Plasma Physics Experiments, World Scientific Publications, London, p. 80 (1988).

25. M. Milanese, R. Moroso, J. Pouzo, "Plasma Filamentation and Upper Pressure Limit for Neutron Yield in a DPF Device", IEEE Transactions on Plasma Science (1993), Vol. 21, No. 5, p. 606.

26. M. Milanese, J. Pouzo, "Neutron Yield Scaling Laws for Plasma Focus Devices", Small Plasma Physics Experiments, World Scientific Publications, London, p. 66 (1988).

27. H Bruzzone, H Acu˜na, M Barbaglia and A Clausse, "A simple plasma diagnostic based on processing the electrical signals from coaxial discharges" Plasma Phys. Control. Fusion 48 (2006) 609–620.

28. C.R. Haas, R. Noll, F. Rohl, G. Herziger, "Schlieren diagnostics of the plasma focus" Nucl. Fusion 24 (1984) 1216-1220.

29. V A Gribkov, A Banaszak, B Bienkowska, A V Dubrovsky, I Ivanova-Stanik, L Jakubowski, L Karpinski, R A Miklaszewski, M Paduch, M J Sadowski, M Scholz, A Szydlowski and K Tomaszewski, "Plasma dynamics in the PF-1000 device under full-scale energy storage: II. Fast electron and ion characteristics versus neutron emission parameters and gun optimization perspectives" J. Phys. D: Appl. Phys. 40 (2007) 3592–3607.

30. http://www.intimal.edu.my/school/fas/UFLF/machines/

31. M. Scholz, B. Bieńkowska, M. Borowiecki, I. Ivanova-Stanik, L. Karpiński, W. Stępniewski, M. Paduch, K. Tomaszewski, M. Sadowski, A. Szydłowski, P. Kubeš, J. Kravárik, "Status of a mega-joule scale Plasma-Focus experiments", *NUKLEONIKA*, vol 51, pp 79-84 ,2006.

32. S. K. H. Auluck "New terms in MHD equations and their impact on the inertial confinement fusion concept", J. Plasma Phys. 36 (1986) p. 211

33. Eric J. Lerner, S. Krupakar Murali, Aaron M. Blake, Derek M. Shannon, Frederik J. van Roessel,"Fusion reaction scaling in a mega-amp dense plasma focus", NUKLEONIKA ;57(2): 205−209, (2012).

34. Sing Lee, Paul Lee, Guixin Zhang, Xianping Feng, V. A. Gribkov, Mahe Liu, Adrian Serban, and Terence K. S. Wong, "High Rep Rate High Performance Plasma Focus as a Powerful Radiation Source" IEEE TRAN. PLASMA SCIENCE, 26, (1998), pp 1119-1126.

Appendix

The metric for the local coordinate system can be written as

$$ds^2 = d\xi^2 + \tilde{r}^2(\xi,\zeta)d\theta^2 + d\zeta^2 \qquad \text{A.1}$$

The scale factors are then $h_1 = 1, h_2 = \tilde{r}(\xi,\zeta), h_3 = 1$. Differential operators can then be written as

$$\vec{\nabla}F = \hat{\xi}\partial_\xi F + \hat{\theta}\tilde{r}^{-1}(\xi,\zeta)\partial_\theta F + \hat{\zeta}\partial_\zeta F$$

$$\vec{\nabla}\cdot\vec{V} \equiv \frac{1}{\tilde{r}(\xi,\zeta)}\left\{\frac{\partial(\tilde{r}(\xi,\zeta)V_\xi)}{\partial\xi} + \frac{\partial V_\theta}{\partial\theta} + \frac{\partial(\tilde{r}(\xi,\zeta)V_\zeta)}{\partial\zeta}\right\} \qquad \text{A.2}$$

The differentiation of unit vectors is given below:

$$\partial_\xi\hat{\xi} = \hat{\zeta}\mathbb{C}(\xi,\zeta)\,;\,\partial_\xi\hat{\theta} = 0\,;\,\partial_\xi\hat{\zeta} = -\hat{\zeta}\mathbb{C}(\xi,\zeta)\,;\,\mathbb{C}(\xi,\zeta) \equiv s\tilde{r}^{-2}(\xi,\zeta)\left(\xi\partial_\xi N(\xi,\zeta) - N(\xi,\zeta)\right)$$

$$\partial_\theta\hat{\xi} = \tilde{r}^{-1}(\xi,\zeta)\xi\hat{\theta}\,;\;\partial_\theta\hat{\theta} = -\hat{\xi}\tilde{r}^{-1}(\xi,\zeta)\xi + \tilde{r}^{-1}(\xi,\zeta)sN(\xi,\zeta)\hat{\zeta}\,;\partial_\theta\hat{\zeta} = -\hat{\theta}\tilde{r}^{-1}(\xi,\zeta)sN(\xi,\zeta) \qquad \text{A.3}$$

$$\partial_\zeta \hat{\xi} = \hat{\zeta}\mathbb{D}(\xi,\zeta); \partial_\zeta \hat{\theta} = 0; \partial_\zeta \hat{\zeta} = -\hat{\xi}\mathbb{D}(\xi,\zeta); \mathbb{D}(\xi,\zeta) \equiv \tilde{r}^{-2}(\xi,\zeta)\xi s \partial_\zeta N(\xi,\zeta)$$

The convective acceleration becomes

$$(\vec{v}\cdot\vec{\nabla})\vec{v} = \hat{\xi}\left\{v_\xi \partial_\xi v_\xi + v_\theta \tilde{r}^{-1}(\xi,\zeta)\partial_\theta v_\xi + v_\zeta \partial_\zeta v_\xi - \frac{\xi v_\theta^2}{\tilde{r}^2(\xi,\zeta)} - v_\zeta^2 \mathbb{D}(\xi,\zeta)\right\}$$

$$+\hat{\theta}\left\{v_\xi \partial_\xi v_\theta + v_\theta \tilde{r}^{-1}(\xi,\zeta)\hat{\theta}\partial_\theta v_\theta + v_\zeta \partial_\zeta v_\theta + \frac{\xi v_\theta v_\xi}{\tilde{r}^2(\xi,\zeta)} - \frac{s v_\theta v_\zeta N(\xi,\zeta)}{\tilde{r}^2(\xi,\zeta)}\right\} \qquad \text{A.4}$$

$$+\hat{\zeta}\left\{\begin{array}{l} v_\xi \partial_\xi v_\zeta + v_\theta \tilde{r}^{-1}(\xi,\zeta)\partial_\theta v_\zeta + v_\zeta \partial_\zeta v_\zeta + \dfrac{v_\theta^2 s N(\xi,\zeta)}{\tilde{r}^2(\xi,\zeta)} + v_\xi v_\xi \mathbb{C}(\xi,\zeta) - v_\xi v_\zeta \mathbb{C}(\xi,\zeta) \\ +v_\zeta v_\xi \mathbb{D}(\xi,\zeta) \end{array}\right\}$$